\documentclass[sigconf,nonacm]{acmart}

\usepackage{amsmath}
\usepackage{booktabs}
\usepackage{multirow}
\usepackage{algorithm}
\usepackage{algorithmic}
\usepackage{xcolor}
\usepackage{tcolorbox}
\tcbuselibrary{listings,skins,breakable}
\usepackage{listings}
\usepackage{xcolor,colortbl}
\usepackage{ulem}
\usepackage{graphicx}
\usepackage{newunicodechar}

\AtBeginDocument{%
  }
    
\setcopyright{none}  
\renewcommand\footnotetextcopyrightpermission[1]{}  
\begin{document}

\title{STAR: Semantic Table Representation with Header-Aware Clustering and Adaptive Weighted Fusion}

\author{Shui-Hsiang Hsu}
\affiliation{%
  \institution{National Chung Hsing University}
  \institution{Smart Sustainable New Agriculture Research Center (SMARTer)}
  \city{Taichung}
  \country{Taiwan}}
\email{g113056055@smail.nchu.edu.tw}

\author{Tsung-Hsiang Chou}
\affiliation{%
  \institution{National Chung Hsing University}
    \institution{Smart Sustainable New Agriculture Research Center (SMARTer)}
  \city{Taichung}
  \country{Taiwan}}
\email{yumeow0122@smail.nchu.edu.tw}

\author{Chen-Jui Yu}
\affiliation{%
  \institution{National Chung Hsing University}
  \institution{Smart Sustainable New Agriculture Research Center (SMARTer)}
  \city{Taichung}
  \country{Taiwan}}
\email{rui0828@smail.nchu.edu.tw}

\author{Yao-Chung Fan}
\affiliation{%
  \institution{National Chung Hsing University}
  \institution{Smart Sustainable New Agriculture Research Center (SMARTer)}
  \city{Taichung}
  \country{Taiwan}}
\email{yfan@nchu.edu.tw}

\begin{abstract}
Table retrieval is the task of retrieving the most relevant tables from large-scale corpora given natural language queries. However, structural and semantic discrepancies between unstructured text and structured tables make embedding alignment particularly challenging.
Recent methods such as QGpT attempt to enrich table semantics by generating synthetic queries, yet they still rely on coarse partial-table sampling and simple fusion strategies, which limit semantic diversity and hinder effective query–table alignment.
We propose STAR (Semantic Table Representation), a lightweight framework that improves semantic table representation through semantic clustering and weighted fusion. STAR first applies header-aware K-means clustering to group semantically similar rows and selects representative centroid instances to construct a diverse partial table.
It then generates cluster-specific synthetic queries to comprehensively cover the table's semantic space.
Finally, STAR employs weighted fusion strategies to integrate table and query embeddings, enabling fine-grained semantic alignment.
This design enables STAR to capture complementary information from structured and textual sources, improving the expressiveness of table representations.
Experiments on five benchmarks show that STAR achieves consistently higher Recall than QGpT on all datasets, demonstrating the effectiveness of semantic clustering and adaptive weighted fusion for robust table representation.
Our code is available at \url{https://github.com/adsl135789/STAR}.
\end{abstract}



\maketitle


\section{Introduction}
\label{sec:introduction}

Table retrieval aims to identify the most relevant tables from a corpus given a natural language query. However, a substantial semantic gap exists between unstructured queries and structured tables \cite{huang2022mixed}, limiting the effectiveness of dense retrieval methods. In practice, serialized tables frequently exceed encoder token limits, making query–table semantic matching more challenging.

To address the semantic gap and long table scenarios, recent studies have attempted to address these issues through semantic augmentation. QGpT \cite{liang-etal-2025-improving-table} adopts a two-stage design: First, it selects the first $k$ rows of the table to form a partial table, which serves as an approximation of the entire table under the token length restriction. Then, it uses an LLM to generate synthetic queries for the partial table, extending the structured content into natural language for easier semantic matching with queries. However, QGpT has two main limitations: (1) Heuristic instance selection: it uses a simple top-$k$ strategy that lacks representativeness for the entire table; (2) Coarse representation fusion: it concatenates the partial table and synthetic queries into a single sequence, making it difficult to control their relative contributions to the final table representation.
\begin{figure*}[t!]
\centering
\includegraphics[width=1.0\textwidth]{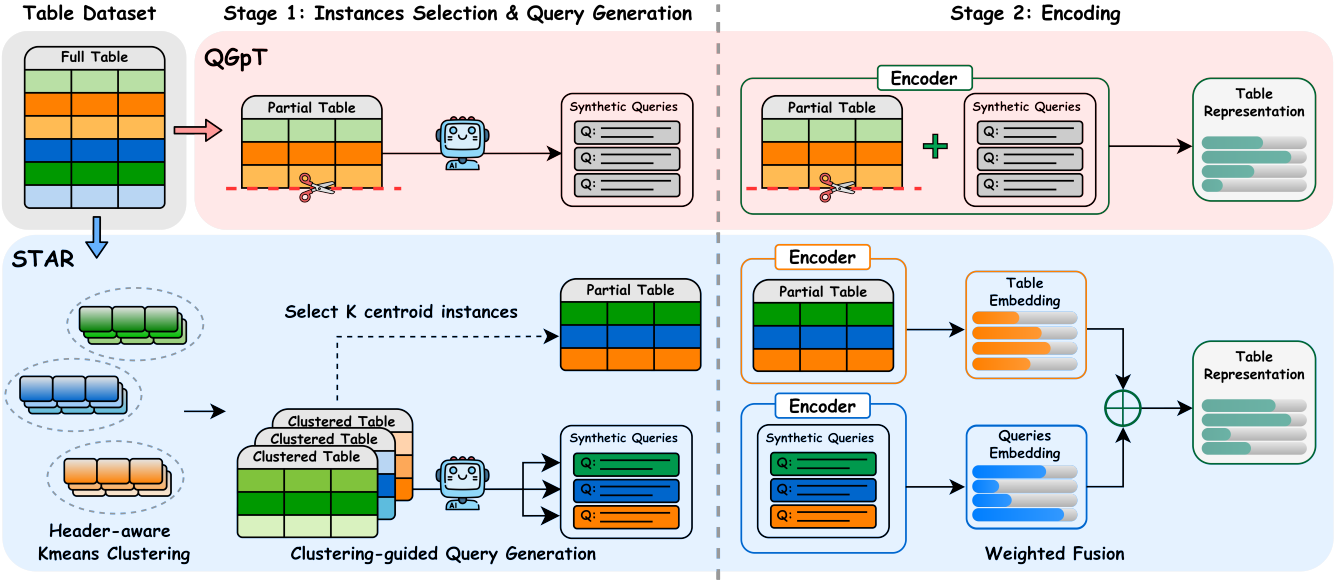}
\caption{Overview of the STAR framework compared to the QGpT baseline. STAR improves representation via two stages: (1) replacing top-$k$ sampling with \textbf{Header-aware Clustering} for diverse instance selection and query generation, and (2) replacing simple concatenation with \textbf{Weighted Fusion} to explicitly model the importance of structured data and synthetic queries. Different colors denote different semantic row clusters.}
\label{fig:comparison}
\end{figure*}

To address these issues, we propose \textbf{STAR} (\textbf{S}emantic \textbf{TA}ble \textbf{R}epresentation), a lightweight framework for table retrieval that improves table representations without modifying the underlying retriever architecture.
STAR captures table semantics by selecting representative instances and augmenting tables with synthetic queries, which are then integrated through a fusion mechanism.
Experimental results on multiple benchmarks demonstrate the effectiveness of the proposed approach. Concretely, our contributions are:
\begin{itemize}
  \item We introduce STAR, a lightweight framework for table retrieval that improves semantic table representations without modifying the retriever architecture.
  \item We propose a clustering-based approach to select representative table content, improving semantic coverage beyond heuristic row sampling.
  \item We present a fusion strategy that integrates tables with synthetic queries for improved query–table alignment.
  \item Experiments on multiple benchmark datasets demonstrate that STAR consistently outperforms baseline methods.
\end{itemize}

\section{Related Works}
\label{sec:related}

Early table retrieval methods primarily relied on sparse retrieval techniques \cite{robertson2009probabilistic}, using lexical matching to compute relevance but struggling to capture deeper semantic relationships. With the advancement of deep learning, dense retrieval methods \cite{karpukhin-etal-2020-dense} leveraging vector representations of text emerged for semantic matching and have gradually been applied to table retrieval tasks. To better handle the structural characteristics of tables, some studies proposed structure-aware encoding methods \cite{herzig2020tapas,yin-etal-2020-tabert}, while recent research has shifted towards query generation for semantic enhancement, using generative models to produce pseudo-queries for improving retrieval performance \cite{mao-etal-2021-generation,wang-etal-2022-gpl}. In the field of table retrieval, QGpT \cite{liang-etal-2025-improving-table} is the first to systematically apply LLMs to generate synthetic queries, enriching table semantic representations.

However, QGpT employs a simple top-k sampling strategy, lacking semantic diversity, and directly concatenates tables and queries, failing to model the importance of different information sources at a fine-grained level. Previous research has also highlighted that heterogeneous signals from tables and text should be modeled separately before fusion, rather than treated as a simple sequence \cite{zayats-etal-2021-representations}. This paper ensures instance diversity through header-aware clustering \cite{jung2025haetae} and introduces a weighted fusion strategy for fine-grained semantic integration.

\section{Methodology}
\label{sec:methodology}

Given a table \(\mathcal{T}\) with a header \(\mathcal{H}\) and \(n\) instances \(\{\mathbf{r}_1, \mathbf{r}_2, \ldots, \mathbf{r}_n\}\), and a user query \(q\), the goal of table retrieval is to retrieve the most relevant tables from a corpus \(\mathcal{C}\) based on \(q\). The STAR framework consists of two stages: (1) Semantic Clustering and Query Generation (SCQG), and (2) Weighted Fusion (WF).

\subsection{Semantic Clustering and Query Generation}

As shown in Figure \ref{fig:comparison} (Stage 1), QGpT adopts a simple top-k strategy to select the top-k instances as the partial table. To address the lack of diversity in instance selection and incomplete query generation, we propose SCQG, which combines semantic clustering with cluster-guided query generation.

\subsubsection{\textbf{Header-aware K-means Clustering.}} Clustering solely on instances ignores the table schema, potentially misgrouping instances with different meanings. We thus integrate header semantics to capture the global context. Specifically, we first use a pretrained encoder to encode the header \(\mathcal{H}\) and each instance \(\mathbf{r}_i\), obtaining the header embedding \(\mathbf{e}_{\mathcal{H}}\) and instance embedding \(\mathbf{e}_{\mathbf{r}_i}\):
\begin{align}
\mathbf{e}_{\mathcal{H}} &= \text{Encoder}(\mathcal{H}) \\
\mathbf{e}_{\mathbf{r}_i} &= \text{Encoder}(\mathbf{r}_i)
\end{align}

We then perform a weighted fusion of the header embedding \(\mathbf{e}_{\mathcal{H}}\) and the instance embeddings \(\mathbf{e}_{\mathbf{r}_i}\) to form the header-aware instance embedding:
\begin{equation}
\mathbf{e}_i = \alpha \cdot \mathbf{e}_{\mathcal{H}} + (1-\alpha) \cdot \mathbf{e}_{\mathbf{r}_i}
\end{equation}
where \(\alpha\) is a hyperparameter controlling the relative importance of the header and the instance, combining both instance semantics and header structural information.

Next, we perform K-means clustering on the embeddings\\ \(\{\mathbf{e}_1, \mathbf{e}_2, \ldots, \mathbf{e}_n\}\) and group them into \(k\) clusters \(\{C_1, C_2, \ldots, C_k\}\). For each cluster \(C_j\), we select the instance closest to the cluster center as the representative:
\begin{equation}
\mathbf{r}_j^* = \arg\min_{\mathbf{r}_i \in C_j} \|\mathbf{e}_i - \boldsymbol{\mu}_j\|^2
\end{equation}
where \(\boldsymbol{\mu}_j\) is the center of cluster \(C_j\). The resulting partial table \(\mathcal{T}_{partial} = \{\mathbf{r}_1^*, \mathbf{r}_2^*, \ldots, \mathbf{r}_k^*\}\) preserves both semantic diversity and representativeness.

\subsubsection{\textbf{Clustering-Guided Query Generation.}}
For each cluster \(C_j\), we construct a clustered table \(\mathcal{T}_j = \{\mathcal{H}, \mathbf{r} \mid \mathbf{r} \in C_j\}\) and generate a synthetic query \(q_j\) using a large language model (LLM):
\begin{equation}
q_j = \text{LLM}(\text{Prompt}(\mathcal{T}_j))
\end{equation}

Thus, for each table \(\mathcal{T}\), we generate \(k\) synthetic queries\\\(\{q_1, q_2, \ldots, q_k\}\), each corresponding to a specific semantic cluster. Compared to QGpT's top-k sampling strategy for generating queries from a partial table, our cluster-specific generation ensures that the queries cover the diverse aspects represented by the table's different semantic clusters, acting as a semantic bridge between the user query and the table.

\begin{table*}[t]
\centering
\caption{Performance comparison on five datasets with different fusion strategies. Bold: best, underlined: second-best.}
\label{tab:main_results}
\resizebox{\textwidth}{!}{%
\begin{tabular}{l*{18}{c}}
\toprule
\textbf{Method} &
\multicolumn{3}{c}{\textbf{Mimo (ch)}} &
\multicolumn{3}{c}{\textbf{Mimo (en)}} &
\multicolumn{3}{c}{\textbf{OTTQA}} &
\multicolumn{3}{c}{\textbf{FetaQA}} &
\multicolumn{3}{c}{\textbf{E2E-WTQ}} &
\multicolumn{3}{c}{\textbf{Avg.}} \\
\cmidrule(lr){2-4} \cmidrule(lr){5-7} \cmidrule(lr){8-10}
\cmidrule(lr){11-13} \cmidrule(lr){14-16} \cmidrule(lr){17-19}
& R@1 & R@5 & R@10 & R@1 & R@5 & R@10 & R@1 & R@5 & R@10
& R@1 & R@5 & R@10 & R@1 & R@5 & R@10 & R@1 & R@5 & R@10 \\
\midrule
QGpT
& 49.81 & 71.06 & 77.23
& 50.66 & 72.35 & 80.80
& 51.45 & 78.14 & 86.68
& 33.95 & 50.87 & 57.86
& 41.49 & 65.98 & 72.61
& 45.47 & 67.88 & 75.04 \\
\midrule
\multicolumn{19}{l}{STAR w/ FWF ($\lambda:$ weight of query embeddings)} \\
\quad $\lambda = 0.1$
& 47.14 & 69.06 & 76.99
& 53.56 & 74.43 & 80.93
& 52.21 & 78.77 & 86.86
& 34.55 & 53.37 & 62.01
& 55.60 & \uline{85.48} & 89.63
& 48.61 & 72.22 & 79.28 \\
\quad $\lambda = 0.2$
& 49.58 & 71.79 & 77.54
& 56.89 & 76.14 & 81.92
& 53.52 & \uline{80.08} & 87.53
& 35.40 & 54.22 & \textbf{62.51}
& \uline{56.85} & 85.06 & 89.63
& 50.45 & 73.46 & 79.83 \\
\quad $\lambda = 0.3$
& 51.36 & \textbf{72.16} & \textbf{78.08}
& 58.40 & 76.98 & 82.50
& 53.84 & \textbf{80.17} & \uline{88.17}
& \uline{36.00} & \uline{54.92} & \uline{62.21}
& \textbf{58.51} & \textbf{85.89} & \uline{90.04}
& \uline{51.62} & \textbf{74.02} & 80.20 \\
\quad $\lambda = 0.4$
& \textbf{52.26} & 71.95 & 77.73
& 58.79 & \uline{77.56} & 82.66
& \uline{53.88} & 79.95 & \textbf{88.53}
& 35.90 & \textbf{55.02} & 62.11
& \uline{56.85} & 85.06 & \uline{90.04}
& 51.54 & 73.91 & \uline{80.21} \\
\quad $\lambda = 0.5$
& 51.56 & 71.49 & 76.82
& \textbf{59.10} & 77.30 & \textbf{83.26}
& 53.79 & 79.81 & 87.58
& 35.75 & 54.37 & 61.76
& 56.43 & 83.40 & 89.63
& 51.33 & 73.27 & 79.81 \\
\quad $\lambda = 0.6$
& 50.18 & 71.26 & 77.39
& 58.79 & 77.17 & \uline{83.02}
& 53.16 & 78.55 & 86.86
& 34.90 & 54.12 & 61.16
& 54.77 & 82.16 & 89.63
& 50.36 & 72.65 & 79.61 \\
\quad $\lambda = 0.7$
& 48.94 & 69.52 & 77.10
& 56.53 & 76.09 & \uline{83.02}
& 51.90 & 77.60 & 85.46
& 34.10 & 52.92 & 60.51
& 54.77 & 82.16 & 87.14
& 49.25 & 71.66 & 78.65 \\
\quad $\lambda = 0.8$
& 47.12 & 68.66 & 76.26
& 55.75 & 75.00 & 81.34
& 50.32 & 76.15 & 84.28
& 32.65 & 50.72 & 59.11
& 53.94 & 81.33 & 85.89
& 47.96 & 70.37 & 77.38 \\
\quad $\lambda = 0.9$
& 45.32 & 67.40 & 74.84
& 54.97 & 73.21 & 80.56
& 48.46 & 74.25 & 82.70
& 31.60 & 48.88 & 57.51
& 52.70 & 80.08 & 85.48
& 46.61 & 68.76 & 76.22 \\
\midrule
STAR w/ DWF
& \uline{51.58} & \uline{72.15} & \uline{77.99}
& \uline{58.89} & \textbf{77.72} & 82.89
& \textbf{54.07} & 79.99 & 88.08
& \textbf{36.25} & 54.77 & \uline{62.21}
& \textbf{58.51} & 85.06 & \textbf{90.06}
& \textbf{51.86} & \uline{73.94} & \textbf{80.25} \\
\bottomrule
\end{tabular}
}
\end{table*}

\subsection{Weighted Fusion}

As shown in Figure \ref{fig:comparison} (Stage 2), QGpT adopts a simple concatenation strategy, where the partial table and synthetic queries are concatenated and passed into a single encoder to obtain a unified representation. This approach mixes different sources of information without modeling the semantic importance of each component. In contrast, we propose a weighted fusion strategy, where the table and queries are encoded separately and then fused with weights.

Specifically, we first encode the partial table and the concatenation of all synthetic queries:
\begin{align}
\mathbf{e}_{\text{table}} &= \text{Encoder}(\mathcal{T}_{partial}) \\
\mathbf{e}_{\text{queries}} &= \text{Encoder}(q_1 \oplus q_2 \oplus \cdots \oplus q_k)
\end{align}

The final table representation is obtained through weighted fusion:
\begin{equation}
\mathbf{e}_{\mathcal{T}} = w_t \cdot \mathbf{e}_{\text{table}} + w_q \cdot \mathbf{e}_{\text{queries}}
\end{equation}
where \(w_t\) and \(w_q\) are the weights for the table and queries, respectively, and \(w_t + w_q = 1\). 

We propose two weight allocation strategies:
\subsubsection{\textbf{Fixed Weight Fusion (FWF)}}
This strategy assigns a constant scalar \(w_q = \lambda\) to the queries, with the remaining weight allocated to the table: \(w_t = 1 - \lambda\). This method is simple and intuitive, and adjusting \(\lambda\) controls the relative importance of the table and queries in the final representation.
\subsubsection{\textbf{Dynamic Weight Fusion (DWF)}}
In contrast, this method dynamically adapts the weights based on the semantic similarity between the table and queries. We calculate the cosine similarity:
\begin{equation}
s = \cos(\mathbf{e}_{\text{table}}, \mathbf{e}_{\text{queries}})
\end{equation}

The weights are then determined by the similarity:
\begin{equation}
w_q = \beta \cdot s, \quad w_t = 1 - w_q
\end{equation}
where \(\beta\) is a scaling factor that controls the impact of similarity changes on the weights. 

\section{Experiments}
\label{sec:experiments}

\subsection{Experimental Setup}

\subsubsection{\textbf{Datasets}}
We evaluate STAR on five widely used table retrieval benchmark datasets:
(1) \textbf{Mimo (ch)}\cite{li-etal-2025-mimotable}: A Chinese table retrieval dataset;
(2) \textbf{Mimo (en)}\cite{li-etal-2025-mimotable}: An English table retrieval dataset;
(3) \textbf{OTTQA}\cite{chen2021open}: An open-domain table question answering dataset;
(4) \textbf{FetaQA}\cite{nan-etal-2022-fetaqa}: A table-based question answering dataset;
(5) \textbf{E2E-WTQ}\cite{pan2021cltr}: An end-to-end table question answering dataset.

\subsubsection{\textbf{Baselines}}
We compare with QGpT\cite{liang-etal-2025-improving-table}, which generates synthetic queries using LLM for table retrieval.

\subsubsection{\textbf{Implementation}}
To ensure a fair comparison, we follow the original experimental setup of QGpT. We use BGE-M3 as the encoder and Llama 3.1 8B-Instruct to generate synthetic queries. The number of clusters \(k\) is set to 10, so each table generates 10 queries. In header-aware K-means clustering, we set \(\alpha = 0.2\). To explore the impact of different weight configurations, we test multiple values of \(\lambda\) for Fixed Weight Fusion, with \(\lambda \in \{0.1, 0.2, \dots, 0.9\}\); for Dynamic Weight Fusion, we set \(\beta = 0.5\). All experiments are evaluated using Recall@K (\(K \in \{1, 5, 10\}\)) as the evaluation metric.

\subsection{Main Experimental Results}


Table \ref{tab:main_results} presents a comparison of STAR and QGpT across five datasets, along with the performance of different weighted fusion strategies. We can observe the following points:

\paragraph{\textbf{STAR significantly outperforms QGpT}}
The STAR framework outperforms QGpT under all fusion strategies. For example, with DWF, STAR's average R@1 improves by 6.39 percentage points, and R@5 also increases by 6.06 percentage points. This shows that the semantic clustering and weighted fusion strategies proposed in STAR are effective at capturing the semantics of tables and improving retrieval performance.

\paragraph{\textbf{Different weights in FWF reflect dataset characteristics}}
Analyzing the performance of different weights (\(\lambda\)) in FWF, we observe that different datasets have varying preferences for query weight. The Mimo (ch) and Mimo (en) datasets perform best when \(\lambda\) is between 0.4 and 0.5. This is likely because the Mimo tables originate from real-world Excel files, which have higher structural complexity. In this case, synthetic queries serve as a more effective semantic bridge between user queries and complex tables. In contrast, the OTTQA, FetaQA, and E2E-WTQ datasets achieve the best FWF performance at lower values of \(\lambda\) (e.g., 0.3 or 0.4). These datasets contain simpler table structures, where the table's inherent semantics are already clear. Thus, giving higher weight to the table itself leads to better retrieval performance.

\paragraph{\textbf{DWF demonstrates strong adaptability and potential}}
DWF demonstrates strong adaptability and runs robustly across different datasets. DWF performs best on OTTQA, FetaQA, and E2E-WTQ, while also achieving competitive results on the both Mimo datasets. This indicates that DWF can effectively balance the semantic sources of different tables through dynamic weighting without requiring manual tuning. In the future, this could be replaced with a more powerful learnable module for end-to-end adaptive weighting.

\section{Analysis}
\label{sec:analysis}

\subsection{Ablation Study of STAR Components}


To verify the effectiveness of each component of STAR, we conducted ablation experiments. Table \ref{tab:ablation} shows the average Recall metrics across five datasets. Specifically, we evaluate the following four model variants:

\begin{itemize}
  \item \textbf{STAR (full)}: Full model with SCQG and Dynamic Weight Fusion.
  \item \textbf{w/o SCQG}: SCQG removed; tables are represented using top-$k$ row sampling and direct query generation.
  \item \textbf{w/o WF}: Removing the weighted fusion mechanism and encoding tables and synthetic queries via simple concatenation.
  \item \textbf{w/o Header-aware}: Removing header information from clustering and performing K-means using only instance embeddings.
\end{itemize}

Based on these variants, the experimental results reveal the following observations:

\paragraph{\textbf{Importance of the SCQG stage.}}
Among all components, SCQG contributes the most to overall performance.
Removing SCQG leads to a substantial drop in average R@1 from 51.86\% to 47.07\% ($-4.79$), indicating that semantic clustering plays a critical role in selecting representative instances and enabling diverse, cluster-aware query generation.

\paragraph{\textbf{Effect of Weighted Fusion.}}
Compared to SCQG, weighted fusion provides a complementary but smaller gain.
When replacing weighted fusion with simple concatenation (w/o WF),
the average R@1 decreases by 2.78 points, suggesting that explicitly modeling the relative importance of structured tables and synthetic queries
is beneficial for fine-grained semantic alignment.

\paragraph{\textbf{Role of Header-aware Clustering.}}
Incorporating header semantics yields a consistent improvement.Removing header information from clustering results in a 1.29-point drop in R@1, indicating that header-aware embeddings help stabilize clustering quality by providing global schema context, even when instance semantics dominate.

\begin{table}[t]
\centering
\caption{Ablation Study of STAR Components (avg. across five datasets).}
\label{tab:ablation}
\small
\renewcommand{\arraystretch}{1.1}
\begin{tabular}{l@{\hskip 10pt}c@{\hskip 10pt}c@{\hskip 10pt}c}
\toprule
\textbf{Method} & \textbf{R@1} & \textbf{R@5} & \textbf{R@10} \\
\midrule
STAR (full) & \textbf{51.86} & \textbf{73.94} & \textbf{80.25} \\
\midrule
\quad w/o SCQG & 47.07 {\scriptsize\color{red}(-4.79)} & 68.43 {\scriptsize\color{red}(-5.51)} & 75.88 {\scriptsize\color{red}(-4.37)} \\
\quad w/o WF & 49.08 {\scriptsize\color{red}(-2.78)} & 69.69 {\scriptsize\color{red}(-4.25)} & 77.02 {\scriptsize\color{red}(-3.23)} \\
\quad w/o Header-aware & 50.57 {\scriptsize\color{red}(-1.29)} & 73.63 {\scriptsize\color{red}(-0.31)} & 79.92 {\scriptsize\color{red}(-0.33)} \\
\bottomrule
\end{tabular}
\end{table}

\section{Conclusion}
\label{sec:conclusion}

This paper addresses the heuristic sampling and coarse fusion issues in existing table retrieval methods by proposing the STAR framework.
STAR improves table representations through two stages: Semantic Clustering and Query Generation, which applies header-aware K-means clustering to select diverse instances and generate cluster-specific synthetic queries, and Weighted Fusion, which enables fine-grained integration of structured tables and synthetic queries.
Experiments on five benchmark datasets show that STAR consistently outperforms QGpT, achieving an average R@1 improvement of 6.39 percentage points across all datasets.
These results demonstrate the effectiveness of combining semantic clustering with adaptive fusion for robust table representation.

Despite its consistent improvements, STAR has several limitations.
First, the clustering-based design relies on informative table headers; tables with vague or missing headers may benefit less from header-aware clustering.
Second, generating cluster-specific synthetic queries introduces additional computational overhead compared to QGpT, which may affect efficiency in large-scale retrieval settings.

Future work will explore more efficient instance selection strategies, such as lightweight clustering or adaptive row sampling, to reduce computational cost.
In addition, we plan to investigate end-to-end learnable weighting mechanisms to automatically optimize fusion weights based on table and query characteristics.

\begin{acks}
This research was supported (in part) by NSTC 114-2634-F-005-002 - project Smart Sustainable New Agriculture Research Center (SMARTer).
\end{acks}

\bibliographystyle{ACM-Reference-Format}
\bibliography{references}

\appendix
\section{Prompt for Pseudo Query Generation }
\label{sec:appendix}

The following prompt is used to generate a pseudo query for each clustered table. For each cluster \(C_j\), the clustered table \(\mathcal{T}_j\) is formatted and passed to the LLM with this prompt to generate one synthetic query \(q_j\).

\begin{tcolorbox}[
  colback=gray!5,
  colframe=gray!75!black,
  fonttitle=\bfseries,
  title=Prompt,
  enhanced,
  breakable,
  left=3mm,
  right=3mm,
  top=2mm,
  bottom=2mm
]
\small
\ttfamily
You are an expert at analyzing tables and generating a diverse, natural query that could be answered using the table data.\\[0.5em]
\#\#\# Input Table:\\
\{clustered\_table\}\\[0.5em]
\#\#\# Your Task:\\
Generate **one query** based on the actual content and structure of this table.\\[0.5em]
\#\#\# Query Types (choose the most appropriate):\\
- Numerical: "What is the average 'Sales' in 'Region' X?"\\
- List: "List all 'Products' with 'Price' above 100"\\
- Count: "How many 'Orders' have 'Status' = shipped?"\\
- Select: "Which 'Employee' has the highest 'Revenue'?"\\[0.5em]
\#\#\# Important Requirements:\\
- Use **natural, conversational language**\\
- Make the query **specific to the actual content** in the table\\
- Reference real values, names, or entities that appear in the table when possible\\
- For fact-verification style tables, focus on entity-specific and temporal queries\\
- For reasoning-oriented tables, include multi-step or conditional queries\\
- **Language code: \{lang\}**: Generate the query in this language\\[0.5em]
\#\#\# Output Format (JSON only):\\
\texttt{\{\,"query": "your query here"\,\}}\\[0.5em]
Generate the query now:
\end{tcolorbox}

\end{document}